\documentclass[preprint,aps,preprintnumbers]{revtex4}
\usepackage{graphicx,comment,amssymb,latexsym,amsbsy}

\begin{document}

\markboth{Aurel Bulgac, Joaqu\'\i n E. Drut, Piotr Magierski}
{Spin 1/2 Fermions in the Unitary Regime at Finite Temperature}

\title{SPIN 1/2 FERMIONS IN THE UNITARY REGIME AT FINITE TEMPERATURE}
\author{AUREL BULGAC, JOAQUIN E. DRUT}
\address{Department of Physics, University of Washington, Seattle, WA 98195-1560,
USA
}
\author{PIOTR MAGIERSKI}
\address{Faculty of Physics, Warsaw University of Technology, ul. Koszykowa 75, 00-662 Warsaw,
POLAND
}
\begin{abstract}
We have performed a fully non-perturbative calculation of the thermal properties of a system of spin 1/2 fermions in 3D in the unitary regime. We have determined the critical temperature for the superfluid-normal phase transition. The thermodynamic behavior of this system presents a number of unexpected features, and we conclude that spin 1/2 fermions in the BCS-BEC crossover should be classified as a new type of superfluid. 
\end{abstract}
\maketitle
\section{Introduction}

The unitary regime is the situation in which the scattering length $a$ is 
much larger than the average inter-particle distance: $n|a|^3 \gg 1$, where
$n$ is the number density\cite{gfb,ho}. At zero temperature, systems in this 
regime are widely believed to be superfluid, with a coherence length and an 
inter-particle distance of comparable magnitude. Such zero temperature problem 
has been considered by several authors\cite{baker,carlsonchang,giorgini}. It was shown experimentally in 2002 that these systems are (meta)stable, and they have been extensively studied ever since\cite{exp}.
From the theoretical point of view, the typical treatment is based on an idea put
forward by Eagles, Leggett and others\cite{leggett}. Their approach assumes a BCS-like form
for the many-body wave function, which is then used for all values of $a$. 
The main problem with this treatment is that, close to the unitary regime, the fraction of 
non-condensed pairs becomes of order one\cite{stefano}, and so a mean field
description becomes questionable (even including fluctuations).

To determine the thermal properties of fermions in the unitary regime, we have placed 
the system on a 3D cubic spatial lattice, with periodic boundary conditions. 
Since the system under consideration is dilute, the interaction that captures the 
physical situation is a zero-range two-body interaction 
$V(\mathbf{r}_1 - \mathbf{r}_2) = -g \delta(\mathbf{r}_1 - \mathbf{r}_2)$, 
with a momentum cut-off $\hbar k_c$. Details of the method can be found in Ref.~\cite{bdm}.

\section{Results and Conclusions}

\begin{figure}[b]
\begin{tabular}{ccccc}
\includegraphics[width=10cm]{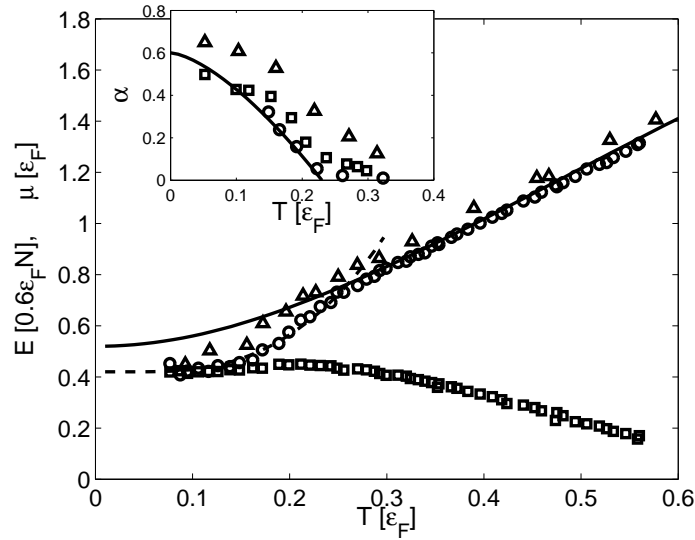}
\end{tabular}
\caption{\label{fig:fig1} The total energy $E(T)$ is shown with open circles 
for a $8^3$-lattice and triangles for a $6^3$-lattice. The chemical 
potential $\mu(T)$ shown with squares for the $8^3$-lattice. The 
Bogoliubov-Anderson phonon and fermion quasiparticle contributions 
$E_{ph+qp}(T)$ 
are shown as a dashed line. The solid line
is $E_{Fg}(T)-0.6\varepsilon_FN (1-\xi_n)$, where $E_{Fg}(T)$ is the energy of a
free Fermi gas. 
Upper left inset: condensate fraction
$\alpha(T)$
, with circles for $10^3$, squares for $8^3$ 
and triangles the $6^3$ lattices respectively. The solid curve is 
$\alpha(T) = \alpha(0) [ 1-(T/T_c)^{3/2}]$.}
\end{figure}

The results of our simulations are partially summarized in Fig. 1 (see Ref.~\cite{bdm} for further details). The thermodynamic quantities we computed present a number of features that can be easily identified. 
First, as $T\rightarrow0$ the energy tends to the $T = 0$ results obtained by 
other groups\cite{carlsonchang,giorgini}. This confirms those results, as the algorithms they used suffer from a sign problem, which is not the case in the present approach. Second, there is a low temperature regime and a high temperature regime, separated by what we argue is the critical temperature for the onset of superfluidity, which we estimated to be $T_c = 0.23(2)$.

For $T<T_c$, the T-dependence of the energy can be accounted for by the two types of elementary excitation expected for this system: boson-like Bogoliubov-Anderson phonons and fermion-like gapped Bogoliubov quasiparticles. Their contributions can be estimated assuming that the system is a Fermi superfluid at $T=0$, with a compressibility and pairing gap as determined from results in Ref.~\cite{carlsonchang}. 

The chemical potential $\mu$ is essentially constant for $T<T_c$, a fact reminiscent of the behavior of an ideal Bose gas in the condensed phase, even though we know the system is strongly interacting and superfluid. This unexpected result implies lack of fermionic degrees of freedom at those temperatures.

Universality of the unitary regime together with $\mu(T) = const.$ for $T<T_c$ implies that
\begin{equation}
E(T) = N \frac{3}{5}\epsilon_F\xi\left(\frac{T}{\epsilon_F}\right), \ \ \ \ \  \xi\left(\frac{T}{\epsilon_F}\right) 
= \xi_s + \zeta \left(\frac{T}{\epsilon_F}\right)^n, \ \ \ \ \ n = \frac{5}{2}
\end{equation}
which is the temperature dependence of an ideal Bose condensed gas. According to our results $n = 5/2$ to about $10 \%$.

Above $T_c$ the system is expected to become normal. The energy behaves like the energy of an ideal Fermi gas, plotted in Fig. 1 with a vertical offset. This is surprising, because the estimated pair-breaking temperature is $T^* \simeq 0.55\epsilon_F$  (see Refs.~\cite{Perali,leggett}), implying that for $T_c < T < T^*$ there should be a noticeable fraction of non-condensed pairs. Our results show no hint of their presence in this temperature interval.

The condensate fraction $\alpha(T)$, as defined in Ref.~\cite{stefano}, and evaluated at $r = L/2$ pair separation, is shown in the inset of Fig. 1. $\alpha(T)$ defines the off-diagonal long range order of the two-body density matrix\cite{yang}. The temperature dependence of $\alpha(T)$ is again consistent with $T_c = 0.23(2)$. Moreover, as $T \rightarrow 0$ we recover the $T=0$ results in Ref.~\cite{stefano}. The $T$-dependence of $\alpha(T)$ resembles that of an ideal Bose gas, which comes as a surprise as well.

\begin{figure}[b]
\begin{tabular}{ccccc}
\includegraphics[width=10cm]{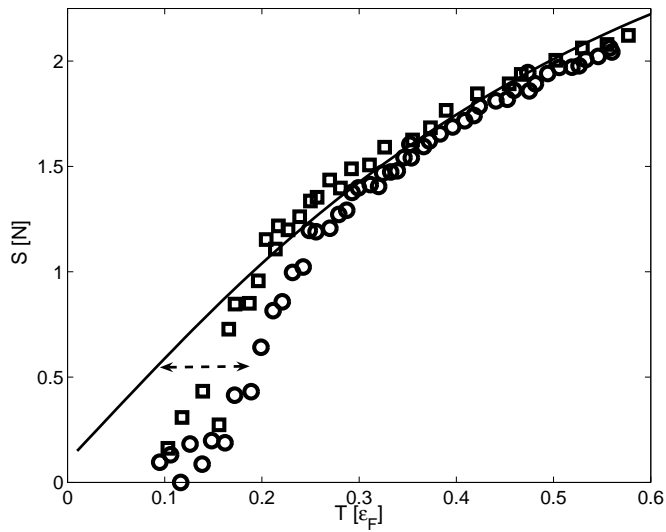}
\end{tabular}

\caption{\label{fig:fig2} Entropy per particle as a function of temperature. Our data in circles for the $8^3$ lattice and in squares for the $6^3$ lattice; free Fermi gas in full line (corresponding to extreme BCS limit). Arrows indicate possible adiabatic cooling or heating processes.}
\end{figure}

At resonance, universality allows us to estimate the entropy as $S(T) = (\frac{5}{3}E(T) -\mu(T) N(T))/T$, see Fig. 2. 
The knowledge of $S(T)$ allows to establish a temperature scale at unitarity. Extending the suggestion of Ref.~\cite{carr}, from known $T$ in the BCS limit, one determines the corresponding $S(T_{BCS})$. By adiabatically tuning the system to the unitary regime one can use $S(T_{BCS}) = S(T_{unitary})$ to determine $T$ in the unitary regime. Especially at very low temperature, our results show relatively large errors in $S$ due to both statistical and finite size effects.


Summarizing, we performed a fully non-perturbative calculation of the thermal properties of a system of spin 1/2 fermions at unitarity. We determined the critical temperature for superfluidity to be $T_c = 0.23(2)$. The thermodynamic behavior of this system presents a number of unexpected features, suggesting that spin 1/2 fermions in the BCS-BEC crossover qualify as a new type of superfluid.


\section{Acknowledgements}

This work was supported by the Department of Energy under grant 
DE-FG03-97ER41014, and by the Polish Committee for Scientific 
Research (KBN) under Contract No. 1 P03B 059 27. 
Use of computers at the Interdisciplinary Centre for Mathematical and 
Computational Modelling (ICM) at Warsaw University is gratefully 
acknowledged. 

\end{document}